\documentclass{llncs}
\usepackage{amsmath}
\usepackage{amssymb}
\usepackage{mathtools}
\usepackage{xspace}
\usepackage{tikz}
\usepackage{fixme}
\usepackage{wrapfig}
\usepackage{paralist,mdwlist}
\usepackage{manfnt,mparhack}
\usepackage[colorlinks=true, citecolor=blue,linkcolor=red,urlcolor=black]{hyperref}
\usepackage{multirow}
\usepackage{framed}
\usepackage{fullpage}
\usepackage{array}

\newcommand{\states}{\ensuremath{S} }
\newcommand{\state}{\ensuremath{s} }
\newcommand{\statesOne}{\ensuremath{S_{1}} }

\newcommand{\supp}{\ensuremath{{\sf Supp}} }
\newcommand{\initState}{\ensuremath{s_{{\sf init}}} }
\newcommand{\edges}{\ensuremath{E} }

\newcommand{\weight}{\ensuremath{w} }

\newcommand{\integ}{\ensuremath{\mathbb{Z}} }
\newcommand{\strat}{\ensuremath{\sigma} }
\newcommand*{\pr}{\mathbb{P}}

\newcommand*{\expect}{\mathbb{E}}

\newcommand{\truncatedTarget}{\ensuremath{T} }
\newcommand{\truncatedSum}[1]{\ensuremath{{\sf TS}^{#1}} }

\newcommand{\dimension}{\ensuremath{d} }

\newcommand{\nat}{\ensuremath{\mathbb{N}} }
\newcommand{\rat}{\ensuremath{\mathbb{Q}} }

\usetikzlibrary{shapes,arrows}
\newcommand{\statesTwo}{\states_2}
\newcommand{\play}{\pi}

\newcommand\calD{\ensuremath{\mathcal{D}}}

\newcommand{\event}{{\cal E}}

\usetikzlibrary{automata,calc}
\usetikzlibrary{arrows}

\title{{\bf Reconciling Rationality and Stochasticity:\\Rich Behavioral Models in Two-Player Games}}
\author{Mickael Randour\inst{1}$^{,}$\thanks{The author is an F.R.S.-FNRS Postdoctoral Researcher}}

\institute{Computer Science Department, Universit\'e Libre de Bruxelles (ULB), Belgium}

\begin{document}
\maketitle
\begin{abstract}
Two traditional paradigms are often used to describe the behavior of agents in multi-agent complex systems. In the first one, agents are considered to be fully rational and systems are seen as multi-player games. In the second one, agents are considered to be fully stochastic processes and the system itself is seen as a large stochastic process. From the standpoint of a particular agent\,---\,having to choose a strategy, the choice of the paradigm is crucial: the most adequate strategy depends on the assumptions made on the other agents.

In this paper, we focus on two-player games and their application to the automated synthesis of reliable controllers for reactive systems\,---\,a field at the crossroads between computer science and mathematics. In this setting, the reactive system to control is a player, and its environment is its opponent, usually assumed to be fully antagonistic or fully stochastic. We illustrate several recent developments aiming to breach this narrow taxonomy by providing formal concepts and mathematical frameworks to reason about richer behavioral models.

The interest of such models is not limited to reactive system synthesis but extends to other application fields of game theory. The goal of our contribution is to give a high-level presentation of key concepts and applications, aimed at a broad audience. To achieve this goal, we illustrate those rich behavioral models on a classical challenge of the everyday life: planning a journey in an uncertain environment.
\end{abstract}

\section{Introduction}
\label{sec:intro}

\smallskip\noindent\textbf{Rationality.} Ever since the seminal book of von Neumann and Morgenstern~\cite{von1944theory}, game theory has grown to be a prominent mathematical framework providing general and powerful concepts and methods to reason about interactions between cooperating and/or competing
agents in complex systems. Applications of game theory are common in diverse fields such as computer science, mathematics, economics, biology, etc.

One of the core underlying concepts of game theory is the \textit{rationality hypothesis}~\cite{osborne1994course}: roughly speaking, agents are modeled as players who have clear personal objectives, are aware of their alternatives, form sound expectations about any unknowns, and choose their actions coherently (regarding some notion of optimality with respect to the situation and the objective). In some sense, players are often seen as selfish beings who place their personal benefit above the common good. This indeed led to a whole area of research where mechanisms are sought to enforce that personal benefits coincide with the common good, thus giving a natural incentive to players to behave in a not-so-selfish way. In the particular setting of zero-sum games, this hypothesis generally results in antagonistic interactions between the players. 

\smallskip\noindent\textbf{Rationality in computer science.} While the very notion of agent rationality is often debated in application fields such as social sciences and economics, there are areas where it is without any doubt a natural hypothesis. One of such areas is our core research field of computer science. Indeed, computer processes for example are fully rational agents: they behave according to a preset set of rules in order to achieve a well-formalized objective, and interact in a usually predictable fashion. Similarly, it is quite common to model their interaction through zero-sum games: e.g., processes competing for access to a shared resource. A very promising application of game theory linked to computer science is the emerging field of \textit{reactive system synthesis}. In a nutshell, computer scientists are envisioning ways to automatically build provably-correct and efficient controllers for so-called reactive systems (i.e., computer systems continuously interacting with an uncontrollable environment). This building process, called \textit{synthesis}, is based on strong mathematical grounds and aims to guarantee the correctness of the synthesized controller: this is notably of tremendous importance for safety-critical systems (power plants, ABS for cars, airplane and railway traffic). At a very high level, the game-theoretic formulation of the synthesis problem would be to consider a two-player zero-sum game between the reactive system to control and the uncontrollable environment. The environment is thus considered to be antagonistic and the goal is to find a strategy for the reactive system that guarantees a correct behavior whatever the actions of the environment. If such a strategy exists, it provides a formal model of the controller to implement. We describe this area in more details and with an illustrative example in Sect.~\ref{sec:synthesis}.

\smallskip\noindent\textbf{Stochasticity.} While game theory is usually focused on rational players, other models have been proposed to model large systems composed of autonomous agents. For example, considering agents as \textit{stochastic processes} is often a sufficient abstraction to reason about macroscopic properties of a complex system with adequate accuracy. When every agent is assumed to be stochastic, the resulting system can itself be described by a stochastic process, for example as a Markov chain~\cite{Puterman-wiley94}. A particularly interesting setting for this paper is the one where a rational agent faces a stochastic one: this can be described as a Markov decision process, combining both controllable and stochastic behaviors~\cite{Puterman-wiley94}. The gap between two-player games (i.e., two rational agents) and Markov decision processes (i.e., one rational agent and one stochastic agent) is filled in stochastic games which combine two rational agents and a stochastic one. Those are also called competitive Markov decision processes in a seminal book by Filar and Vrieze~\cite{filar1997}.

\smallskip\noindent\textbf{The paradigm defines the objective hence the adequate strategy.} Let us take a very simple example to illustrate this. Assume you have to choose a way to get from your home to your workplace: you could take your car, catch a bus, go to the railway station, ride your bike, etc. Maybe you could also take different routes. In this example, you are the rational agent. Now, there is another agent: the environment. Essentially, this agent represents everything that can impact your journey: bad weather, bad traffic, train delay, etc. What assumptions should we make on this agent? If we are considering the average time-to-work, for example when deciding whether to buy a car or a yearly train pass, the natural option would be to see the environment as fully stochastic (following a stochastic model that we obtain from real data for example) and to reason about the \textit{expected time-to-work} as this is what we will observe in the long run. But if we are considering which transportation to take on a particular day, for example the morning before a very important meeting, then the situation is different: in this case, it seems reasonable to choose the option that will \textit{guarantee} that we will not be late, even if this option takes a little more time on average (e.g., traffic can be stopped for a long time because of an accident even if the car is the best option on average). Such a situation is best modeled through a two-player zero-sum game against the environment.

\smallskip\noindent\textbf{New paradigms.} Those two paradigms are the prominent ones in the literature, especially for applications in computer science. But they are not sufficient to reason about more complex situations, where richer behavioral models have to be considered. For example, \textit{one may want to combine a good expected time-to-work with acceptable guarantees on the worst-case reaching-time}. To do so, we need appropriate mathematical frameworks, algorithms and tools to decide the existence of adequate strategies and to exploit them. 

Establishing such meaningful frameworks, both expressive enough to tackle the complexity of real-life applications and abstract enough to allow efficient analysis, is a challenge which is gaining attention in the research community. This paper tries to illustrate, at a very high level and for a large audience, some advances in this area that we have obtained through different collaborations. It is based on a series of papers on the subject~\cite{Ran13,bruyere_STACS2014,DBLP:journals/corr/BruyereFRR14,RRS-cav15,RRS15b,LATA}, most in computer science conferences and journals.

\smallskip\noindent\textbf{Outline.} As stated before, the aim of this contribution is to give a broad picture of several new solution concepts for strategies, mixing assumptions of rationality and stochasticity. The paper only contains the minimal amount of technical details necessary to ensure sound definitions of the different models and we point in each section to the corresponding full papers containing detailed techniques. In Sect.~\ref{sec:synthesis}, we give a comprehensive discussion of the controller synthesis problem and how it can be answered using game theory. To illustrate the approach, a toy example of a lawnmower controller is used as a case study. In Sect.~\ref{sec:rich}, we present the core of the paper: novel solution concepts for rich behavioral models, illustrated through an everyday life application\,---\,the planning of a journey in an uncertain environment. We end the paper with a short conclusion in Sect.~\ref{sec:conclusion}.

\section{Context: controller synthesis for reactive systems}
\label{sec:synthesis}

As mentioned in Sect.~\ref{sec:intro}, our core research field is the automated synthesis of reliable and efficient controllers for reactive systems through game theory. This section is based on~\cite{Ran13}: its goal is to illustrate the cornerstones of such an approach. It is organized as follows: Sect.~\ref{subsec:synthesis} presents the general context in more details; Sect.~\ref{subsec:lawnmower} describes a motivating toy example\,---\,a robotized lawnmower\,---\,in an informal way; Sect.~\ref{subsec:modeling} shows how its interaction with its environment can be formalized through game theory; and Sect.~\ref{subsec:controller} discusses the synthesis process and its outcome on the example. 
\subsection{Automated controller synthesis through game theory}
\label{subsec:synthesis}

Nowadays, more and more aspects of our society depend on \textit{critical reactive computer systems}, i.e., systems that continuously interact with their uncontrollable environment. Think about control programs of power plants, ABS for cars or airplane and railway traffic managing. Therefore, we are in dire need of systems capable of sustaining a safe behavior despite the nefarious effects of their environment.

Good developers know that testing do not capture the whole picture: never will it \textit{proves} that no bug or flaw is present in the considered system. So for critical systems, it is useful to apply \textbf{formal verification}. That means using \textit{mathematical tools} to prove that the system follows a given specification which models desired behaviors. Among those tools are \textit{model checking} techniques, introduced in the 80s~\cite{DBLP:conf/lop/ClarkeE81,DBLP:conf/lics/VardiW86,BK-book08} based on logic, automata and games (seminal works by B\"uchi, Rabin, etc). While verification applies \textit{a posteriori}, checking that the formal model of a system satisfies the needed specification, it is most of the time desirable to start \textit{from} the specification and automatically build a system from it, in such a way that desired properties are proved to be maintained in the process. This \textit{a priori} process is the more ambitious and harder \textbf{synthesis} problem~\cite{church1957applications,ramadge1987supervisory,DBLP:conf/popl/PnueliR89}, on which an important part of the research community is currently focusing its effort.

The mathematical framework we use is \textbf{game theory}~\cite{von1944theory,osborne1994course}. Roughly speaking, we consider a reactive system controller as a player (player 1), and its uncontrollable environment as its adversary (player 2). We model their interactions in a game on a graph~\cite{DBLP:conf/dagstuhl/2001automata}, where vertices model \textit{states} of the system and its environment, and edges model their possible \textit{actions}. Players alternatively decide how to move a pebble on this graph. They choose how to move the pebble according to \textit{strategies}, creating an infinite sequence of states called \textit{play} that represents the \textit{behavior of the system}. The goal of the controller is to enforce a given specification, encoded through a \textit{winning objective} (a play is winning if it belongs to a predefined set of acceptable plays). The goal of the environment, considered antagonistic, is to prevent the controller from enforcing its specification. In our context, establishing \textit{winning strategies} (i.e., strategies that guarantee victory whatever the strategy of the environment) corresponds to synthesizing implementable models of provably correct controllers.

The synthesis process is depicted in Fig.~\ref{fig:process}. In this section, we do not address the full theoretical deepness of such an approach but rather try to motivate and illustrate its usefulness toward an audience who may not be familiar with it, by discussing a motivating toy example.

\tikzstyle{decision} = [diamond, draw,
    text width=4.5em, text badly centered, node distance=3cm, inner sep=0pt]
\tikzstyle{block} = [rectangle, draw, 
    text width=5em, text centered, rounded corners, node distance=3.5cm, minimum height=4em]
\tikzstyle{none} = []
\tikzstyle{line} = [draw, -latex']
\tikzstyle{cloud} = [draw, ellipse, node distance=5cm, text badly centered, minimum height=2em,text width=6em]
    
\begin{figure}[htb]
  \centering
  \resizebox{10cm}{!}{
\begin{tikzpicture}[node distance = 2cm, auto]
    \node [cloud] (sysdesc) {system description};
    \node [cloud, right of=sysdesc] (envdesc) {environment description};
    \node [cloud, right of=envdesc] (infspec) {informal specification};
    \node [block, below right of=sysdesc] (game) {model as a game};
    \node [block, right of=game, node distance=5cm] (objec) {model as winning objectives};
    \node [block, below right of=game] (synth) {synthesis};
    \node [decision, below of=synth] (iswin) {is there a winning strategy ?};
    \node [block, below left of=iswin,text width=8em, node distance=4cm] (revise) {empower system capabilities or weaken specification requirements};
    \node [cloud, below right of=iswin, node distance=4cm] (controller) {strategy\\=\\ controller};
    \path [line,dashed] (sysdesc) |- ([yshift=0.5cm]game.north) -- (game);
    \path [line,dashed] (envdesc) |- ([yshift=0.5cm]game.north) -- (game);
    \path [line,dashed] (infspec) -- (objec);
    \path [line] (game) |- ([yshift=0.5cm]synth.north) -- (synth);
    \path [line] (objec) |- ([yshift=0.5cm]synth.north) -- (synth);
    \path [line] (synth) -- (iswin);
    \path [line] (iswin) -- node [left, xshift=-0.2cm] {no}(revise);
    \path [line] (iswin) -- node {yes}(controller);
\end{tikzpicture}}
      \caption{Controller synthesis through game theory: process.}
\label{fig:process}
  \end{figure}
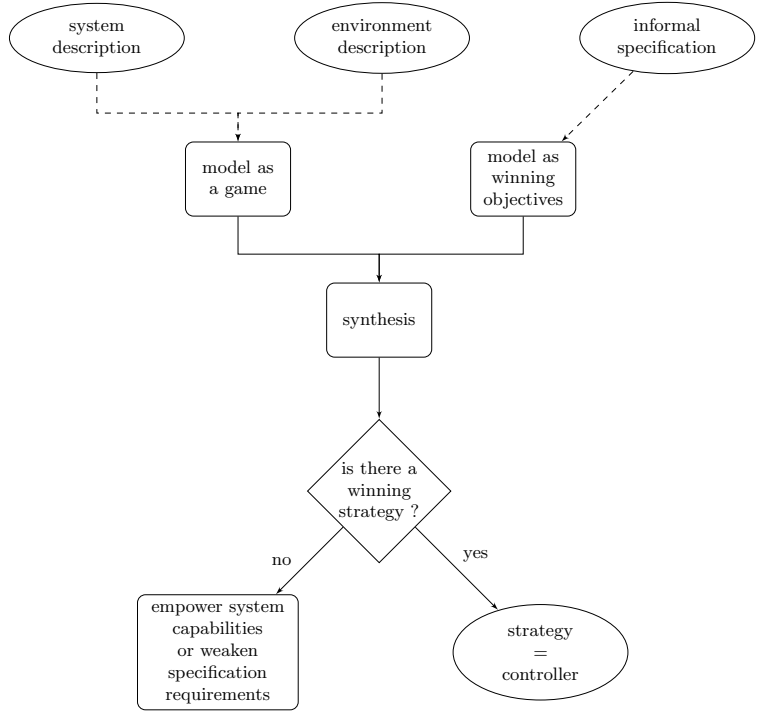

A wide variety of games (and thus system models) have been studied recently, with diverse enforceable behaviors (e.g.,~\cite{DBLP:conf/dagstuhl/2001automata,DBLP:conf/emsoft/ChakrabartiAHS03,DBLP:conf/cav/BloemCHJ09,em79,ZP96,DBLP:journals/iandc/VelnerC0HRR15,DBLP:journals/acta/ChatterjeeRR14,Chatterjee201525}). In our illustration, we focus on systems that must satisfy \textit{qualitative behaviors} (e.g., always eventually granting requests, never reaching a deadlock) along with multiple \textit{quantitative requirements} (e.g., maintaining a bound on the mean response time, never running out of energy). Such settings can be handled using techniques developed by Chatterjee et al.~in~\cite{DBLP:journals/acta/ChatterjeeRR14}.

\subsection{A toy example: the lawnmower}
\label{subsec:lawnmower}

Consider the following running example. We want to synthesize a controller for a robotized lawnmower. This lawnmower is automatically operated, without any human intervention. We present its informal specification, as well as the effects the environment can have on its operation.

\begin{itemize}
\item In this partial, simplified specification, the gardener does not ask for the lawnmower to satisfy any bound on the frequency of grass-cuttings. However, as he wishes that the grass does not grow boundlessly, the lawnmower should cut the grass infinitely often in the future (as if it stops someday, the grass will not stop growing from then on).
\item The lawnmower has an electric battery that can be recharged under sunshine thanks to solar panels, and a fuel tank that can only be filled when the lawnmower is back on its base. Both are considered unbounded to keep things simple.
\item The weather can be cloudy or sunny.
\item The lawnmower can refuel ($2$ fuel units) at its base under both weather conditions, but can only recharge its battery ($2$ battery units) when it is sunny. Resting at the base takes $20$ time units.
\item When cloudy, it can operate either under battery ($1$ battery unit) or using fuel ($2$ fuel units), both according to the same speed ($5$ time units). When sunny, the lawnmower may either cut the grass slowly, which always succeeds and consumes no energy (as the sun recharges the battery along the way), but takes $10$ time units. Or it may cut the grass fast, which consumes both $1$ unit of fuel and $1$ unit of battery, but only takes $2$ time units. 
\item When operating fast, the lawnmower makes considerably much noise, which may wake up the cat that resides in the garden and prompt it to attack the lawnmower. In that case, the grass-cutting is interrupted and the lawnmower goes back to its base, losing $40$ time units as repair is needed. The cat does not go out if the weather is bad.
\item As the gardener cannot benefit from his garden while the lawnmower is operating, he wishes that the mean time required by actions of the lawnmower is less than $10$ time units.
\end{itemize}

While simple, this toy example already involves qualitative requirements (the grass should be mown infinitely often), along with quantitative ones. There are indeed three quantities that have to be taken into account: battery and fuel are energy quantities, which should never be exhausted, and time is a quantity which mean over an infinite operating of the lawnmower should be less than a given bound.

Given this informal description of the capabilities of the system and its environment, as well as the specification the system should enforce, we need to build a system controller that guarantees satisfaction of the specification.

\subsection{Modeling the lawnmower example as a two-player game}
\label{subsec:modeling}

\smallskip\noindent\textbf{Game.} We model the states and the interactions of the couple system/environment as a graph game where the system (here, the lawnmower) is player~1 and the environment is its adversary player~2. Formally, a \textit{game} is a tuple  $G = (\statesOne, \statesTwo, \initState, \edges, \weight)$ where (i) $\statesOne$ and $\statesTwo$ resp.~denote the finite sets of \textit{states} belonging to player~1 and player~2, with $\statesOne \cap \statesTwo = \emptyset$; (ii) $\initState \in \states = \statesOne \cup \statesTwo$ is the initial state; (iii) $\edges \subseteq \states \times \states$ is the set of \textit{edges} such that for all $s \in \states$, there exists $s' \in \states$ such that $(s, s') \in \edges$; and (iv) $\weight\colon \edges \rightarrow \integ^{\dimension}$ is the $d$-dimensional weight labeling function.

The game starts at the initial state $\initState$, and if the current state is a 
player~1 (resp.~player~2) state, then player~1 (resp.~player~2) chooses an outgoing \textit{edge}. This choice is made according to a \textit{strategy} of the player: given the sequence of visited states, a strategy chooses an outgoing edge. For this illustration, we only consider strategies that operate this choice deterministically. This process of choosing edges is repeated forever, and gives rise to an outcome of the game, called a {\em play}, that consists in the infinite sequence of states that are visited. Formally, a \textit{play} in game $G$ is an infinite sequence of states $\play = s_{0}s_{1}s_{2}\ldots{}$ such that $s_{0} = \initState$ and for all $i \geq 0$, we have $(s_{i}, s_{i+1}) \in \edges$.
  
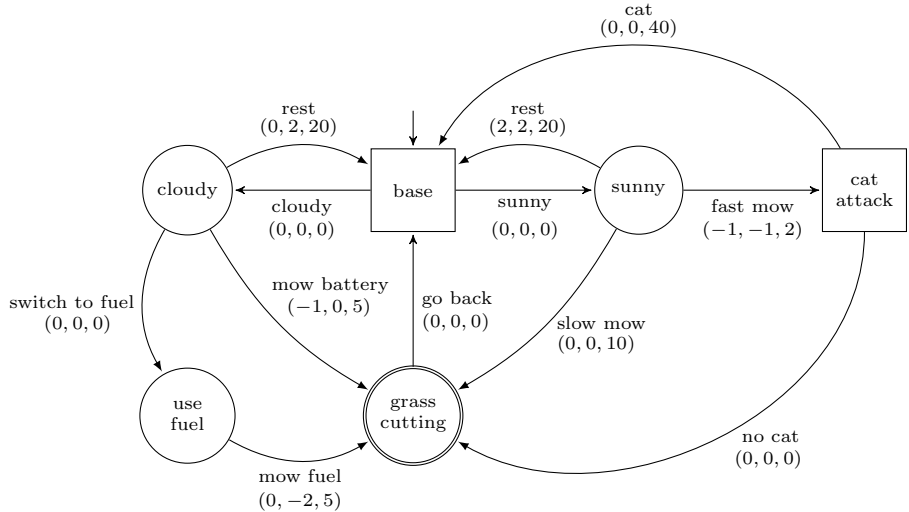
\begin{figure}[bht]
  \centering   
  \begin{tikzpicture}[->,>=stealth',shorten >=1pt,auto,node
    distance=2.5cm,bend angle=45,scale=0.75,font=\scriptsize]
    \tikzstyle{p1}=[draw,circle,text centered,minimum size=11mm, text width=9mm]
    \tikzstyle{p2}=[draw,rectangle,text centered,minimum size=11mm, text width=9mm]
    \node[p1]  (0)  at (0, 0) {cloudy};
    \node[p2]  (1) at (4, 0) {base};
    \node[p1]  (2)  at (8, 0) {sunny};
    \node[p2]  (3)  at (12, 0) {cat attack};
    \node[p1,double]  (4)  at (4, -4) {grass cutting};
    \node[p1]  (5)  at (0, -4) {use fuel};
    \coordinate[shift={(0mm,5mm)}] (init) at (1.north);
    \path
    (1) edge node [below] {cloudy} node [below, yshift=-0.3cm] {$(0,0,0)$} (0)
    (1) edge node [below] {sunny} node [below, yshift=-0.3cm] {$(0,0,0)$}  (2)
    (2) edge node [below] {fast mow} node [below, yshift=-0.3cm] {$(-1,-1,2)$} (3)
    (4) edge node [right] {go back} node [right, yshift=-0.3cm] {$(0,0,0)$} (1)
    (init) edge (1);
	\draw[->,>=latex] (0) to[out=300,in=150] node [right, xshift=-0.2cm, yshift=0.5cm] {mow battery} node [right, yshift=0.2cm] {$(-1,0,5)$} (4);
	\draw[->,>=latex] (0) to[out=240,in=120] node [left] {switch to fuel} node [left, yshift=-0.3cm, xshift=-0.3cm] {$(0,0,0)$} (5);
	\draw[->,>=latex] (5) to[out=330,in=210] node [below] {mow fuel} node [below, yshift=-0.3cm] {$(0,-2,5)$} (4);
	\draw[->,>=latex] (2) to[out=240,in=30] node [right] {slow mow} node [right,yshift=-0.3cm] {$(0,0,10)$} (4);
	\draw[->,>=latex] (3) to[out=270,in=330] node [right,xshift=0.2cm] {no cat} node [right,xshift=0.1cm,yshift=-0.3cm] {$(0,0,0)$} (4);
	\draw[->,>=latex] (0) to[out=30,in=150] node [above,yshift=0.3cm] {rest} node [above] {$(0,2,20)$} (1);
	\draw[->,>=latex] (2) to[out=150,in=30] node [above,yshift=0.3cm] {rest} node [above] {$(2,2,20)$} (1);
	\draw[->,>=latex] (3) to[out=120,in=60] node [above,yshift=0.3cm] {cat} node [above] {$(0,0,40)$} (1);
      \end{tikzpicture}
      \caption{Lawnmower game. Edges are fitted with tuples denoting changes in battery, fuel and time respectively.}
\label{fig:game}
  \end{figure}

Applying this formalism, we represent the lawnmower problem as the game depicted on Fig. \ref{fig:game}. Edges correspond to choices of the system or its environment and taking an edge implies a change on the three considered quantities, as denoted by the edge label. The \textit{grass-cutting} state is special as the specification requires that it should be visited infinitely often by a suitable controller.

\smallskip\noindent\textbf{Strategies.} Formally, a \textit{strategy} for player~$i$, $i \in \lbrace 1, 2\rbrace$, in $G$ is a function $\strat_{i}\colon \states^\ast\states_i \rightarrow \states$ defined over all valid prefixes of plays (w.r.t.~the underlying game graph) and such that for all prefix $\eta = s_{0}s_{1}\ldots{}s_{n}$ with $s_n \in \states_i$, we have $(s_{n}, \strat_{i}(\eta)) \in \edges$. The history of a play (i.e., the previously visited states and their order of appearance) may thus in general be used by a strategy to prescribe its choice. In full generality, strategies may require infinite memory (as they need to recall histories of unbounded length). A strategy $\strat_{i}$ for player~$i$ has \textit{finite memory} if the history it needs to remember can be bounded. In that case, the strategy can be encoded by a specal kind of finite automaton, called deterministic Moore machine. As discussed earlier, a strategy of player~1 (the lawnmower) provides a complete description of a controller for the system, prescribing the actions to take in response to any situation. Therefore, our task is to build a strategy that satisfies the specification.

\smallskip\noindent\textbf{Objectives.} To devise such a strategy, it is needed to formalize the specification as objectives for player~1 in the game. The conjunction of objectives yields a set of winning plays that endorse the specification. A strategy of player~1 is thus said to be \textit{winning} if, \textit{against every possible strategy of the adversary}, the play induced by following this strategy belongs to the winning set of plays.

The informal specification developed in Sect.~\ref{subsec:lawnmower} is encoded as the following objectives. We omit technical details for the sake of this illustration.

\begin{itemize}
\item \textit{Battery and fuel.} Both constitute energy types which quantities are never allowed to drop below zero. A~play is thus winning for the \textit{energy objective} if the running sum of the weights encountered along it (i.e., changes induced by the taken edges) never drops below zero on any of the first two dimensions.
\item \textit{Mean action time.} The specification asks that the lawnmower spends no more than $10$ time units per action on average in the long run. That is, it is allowed to take more than $10$ time units on some actions, but the long-run mean should be below this threshold. Therefore, the \textit{mean-payoff objective} requires that the \textit{limit} of the mean of the third-dimension weights over prefixes of a play is lower than $10$.
\item \textit{Infinitely frequent grass-cutting.} To satisfy this part of the specification, a strategy of player~1 must ensure that the grass-cutting state is visited infinitely often along the induced play. This is encoded as a \textit{B\"uchi objective}.
\end{itemize}

\subsection{Synthesis of a correct and efficient controller}
\label{subsec:controller}

\smallskip\noindent\textbf{Process.} Since in this example, our desire is to build a practical real-world controller, we are only interested in strategies that require \textit{finite} memory. From a theoretical standpoint, there exist classes of games where infinite memory may help to achieve better results (see for example \cite{DBLP:journals/iandc/VelnerC0HRR15,randour2014Thesis}), but infinite-memory strategies are ill-suited for practical use, as implementing a controller with infinite-memory capabilities is ruled out.

The core of the synthesis process depicted on Fig. 1 is thus to construct, if possible, a finite-memory strategy that ensures satisfaction of the previously defined objectives, as well as a corresponding initial value of the energy parameters, commonly referred to as \textit{initial credit}. That is because for the energy objectives, it is allowed to start the game with some finite quantity in stock, before taking any action. Think about starting a race with some fuel in your tank.

While of importance for the analysis of systems with both qualitative and quantitative requirements, the synthesis problem for the class of games that is used to model the lawnmower problem, i.e., games with parity and multi energy or mean-payoff objectives, has only been considered recently~\cite{DBLP:journals/acta/ChatterjeeRR14}. Applying the corresponding techniques, we are able to synthesize a winning strategy for player~1, hence a suitable controller for the lawnmover.

\bigskip\noindent\textbf{Lawnmower controller.} To conclude our illustration, we exhibit a synthesized controller that enforces the desired specification. Notice that there may exist other acceptable controllers. The one we present here is quite simple but already asks for some memory (in the form of bookkeeping of battery and fuel levels). The controller implements the following strategy.
\begin{itemize}
\item Start the game with empty battery and fuel levels.
\item If the weather is sunny, mow slowly.
\item If the weather is cloudy,
\begin{itemize}
\item if there is at least one unit of battery, mow on battery,
\item otherwise, if there is at least two units of fuel, mow on fuel,
\item otherwise, rest at the base.
\end{itemize}
\end{itemize}
Notice that this strategy guarantees never running out of energy (which satisfies the energy objectives), induces infinitely frequent grass-cuttings (which satisfies the B\"uchi objective), and produces a play on which the mean time per action is less than $10$ against any strategy of the adversary (which satisfies the mean-payoff objective). In this sample controller, the lawnmower never uses the ``fast mow'' action as the adversary could very well play ``cat'' and prevent visit of the grass-cutting state.

\subsection{Typical questions and challenges}
The lawnmower illustration is only a toy example, too simple to be of real practical use. Nonetheless, the underlying synthesis techniques aim at dealing with real-world problems, and some of them already proved their interest in industrial contexts. The typical challenges faced in the field are the following.

First, in order to accurately represent the complexity of practical applications, our game models must be \textit{sufficiently expressive}. The needed expressiveness may be in terms of complex winning objectives (e.g.,~\cite{Chatterjee201525,DBLP:journals/corr/BouyerMRLL15}), of the capacity to analyze trade-offs between different objectives (e.g.,~\cite{DBLP:journals/acta/ChatterjeeRR14,RRS-cav15}), of rich interaction between the players~\cite{LATA}, and so on. In Sect.~\ref{sec:rich}, we focus on recent advances toward richer behavioral models~\cite{DBLP:journals/corr/BruyereFRR14,bruyere_STACS2014,RRS15b}.

Second, modeling real applications usually results in huge games, while allowing our techniques to describe complex behavior results in increased computational complexity. For example, it was shown in~\cite{DBLP:journals/acta/ChatterjeeRR14} that synthesizing correct controllers for specifications similar to the lawnmower one requires exponential time in the size of the game. Moreover, optimal controllers for such specifications also require exponential memory. In practice, we favor simpler controllers whenever possible: they are easier to conceive, and cheaper to produce and maintain. Therefore, to preserve the practical interest of synthesis techniques, it is crucial to consider the trade-off between expressiveness and \textit{tractability} of the resulting techniques.

Similarly, it is important that the developed theoretical frameworks and algorithms are then implemented efficiently in \textit{software tools} supporting the synthesis process. This point yields its own challenges, and a lot of effort is put in devising efficient representations and techniques to implement game-theoretic concepts. For example, see~\cite{DBLP:conf/tacas/BohyBFR13} for implementation of techniques linked to the lawnmower case.

\section{Rich behavioral models}
\label{sec:rich}

As argued in Sect.~\ref{sec:intro}, for a rational agent having to adopt a strategy in a multi-agent complex system, the best choice greatly depends on the assumptions made on the other agents. This is especially true in two-player games such as the ones used for synthesis, as presented in Sect.~\ref{sec:synthesis}. Here, we develop the practical case sketched in the introduction: \textit{the dilemma of an employee having to choose a way to get to work}. We show that, depending on the situation, the natural objective of the employee and the assumptions he should make on his environment vary. Then, we illustrate novel concepts that go beyond the two traditional paradigms of fully antagonistic and fully stochastic environments.

This section is based on a more technical survey co-authored with Raskin and Sankur~\cite{RRS15b}. The goal of this section is to illustrate the different concepts intuitively, abstracting most technical details to focus on the philosophy of each solution concept. For each one, we point the interested reader to corresponding publications where the models are formally defined and adequate synthesis techniques are developed. This section is structured as follows:
\begin{itemize}
\item Sect.~\ref{subsec:planning} presents the practical case and defines the necessary notions of the model;
\item Sect.~\ref{subsec:expect} presents a classical solution consisting in minimizing the expected time-to-work, assuming the environment to be fully stochastic;
\item Sect.~\ref{subsec:proba} addresses the case of risk-averse strategies;
\item Sect.~\ref{subsec:wc} shows how to guarantee an acceptable worst-case reaching-time by considering the environment as fully antagonistic;
\item Sect.~\ref{subsec:bwc} presents a novel framework allowing the employee to adopt a strategy ensuring both worst-case guarantees and good expected time-to-work.
\item Finally, Sect.~\ref{subsec:percentile} sketches new concepts that permit reasoning on trade-offs between different aspects: e.g., the employee wants to minimize the risk of being late but also to travel at an acceptable cost.
\end{itemize}

\subsection{Planning a journey in an uncertain environment}
\label{subsec:planning}
~\vspace{-5mm}
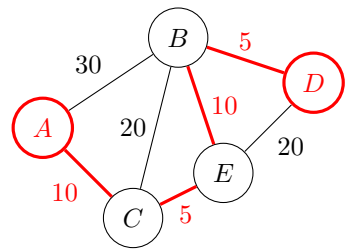
\begin{wrapfigure}{r}{56mm}
\vspace{-5mm}
\centering
\scalebox{1}{\begin{tikzpicture}[-, auto,node
    distance=2.5cm,bend angle=45, scale=0.6, font=\normalsize]
    \tikzstyle{p1}=[draw,circle,text centered,minimum size=7mm,text width=4mm]
    \node[p1,color=red,very thick]  (0)  at (0, 0) {$A$};
    \node[p1]  (1) at (3, 2) {$B$};
    \node[p1]  (2) at (2, -2)  {$C$};
    \node[p1,color=red,very thick]  (3) at (6,1)  {$D$};
    \node[p1]  (4) at (4,-1)  {$E$};
    \path
	(0) edge node[above left] {$30$} (1)
	(0) edge[color=red,very thick] node[below left] {$10$} (2)
	(2) edge node[left] {$20$} (1)
	(2) edge[color=red,very thick] node[below,xshift=1mm] {$5$} (4)
	(4) edge[color=red,very thick] node[right] {$10$} (1)
	(4) edge node[below right] {$20$} (3)
	(1) edge[color=red,very thick] node[above] {$5$} (3);
      \end{tikzpicture}}
  \caption{In a fully deterministic world, traveling would be as easy as finding a shortest path in a graph.}
  \label{fig:graph}
\vspace{-6mm}
\end{wrapfigure}
\smallskip\noindent\textbf{An everyday life challenge.} Let us consider the dilemma of an employee choosing a mean of transport and an itinerary to get to work. This kind of choice is one that we all face at times, some sort of \textit{shortest path problem}: we want to find the quickest way to go from one place to another. The usual way to solve this in the real life is to resort to a GPS navigation device or our favorite web mapping service. Behind the scenes, such a problem is often modeled as finding a path of minimal length in a weighted graph (which may be directed or not). See for example Fig.~\ref{fig:graph}: vertices represent different locations and edges $L \rightarrow L'$ are labeled with a weight that models the time to go from $L$ to $L'$. The problem is then to find a path from the origin ($A$) to the destination ($D$) that minimizes the sum of weights (this path is depicted in thick red in the figure). This is a well-known algorithmic problem with elegant solutions (e.g., Dijkstra and Bellman-Ford algorithms~\cite{cherkassky1996shortest}). But, is it really always that easy? It would be if we lived in a \textit{fully deterministic} world, where going from $A$ to $B$ \textit{always} take the same precise amount of time. But we all know that uncontrollable events can impact a journey: bad traffic can slow our car, trains can be delayed, etc. So, what can we do?

\smallskip\noindent\textbf{Modeling an uncertain environment.} One traditional answer to this observation is to go from a model where the employee is the only agent with choices (a graph) to a model where a second agent represents the environment and interacts with the employee. As discussed earlier, two classical assumptions can be made on the environment: either we see it as a rational agent (yielding a two-player game) or as a stochastic agent (yielding a Markov decision process). In our case, the environment represents traffic, weather, etc. While it may sometimes \textit{seem} like other people are taking the road only to slow us down, it is reasonable to assume that the environment is not a rational agent per se. On the contrary, a stochastic agent is a good model of such phenomena: one can obtain data estimating the probability that a specific part of town is crowded at specific times of the day, the probability that a bus on a given line is delayed, and so on. Hence for the moment, we consider the environment to be \textit{fully stochastic} and the problem to solve becomes a \textit{stochastic shortest path problem}~\cite{Puterman-wiley94}. Still, the very notion of \textit{rationality should not be ruled out} for the environment: we will see that even if the environment is inherently stochastic, the employee sometimes has to see it as a rational, antagonistic agent in order to choose adequate strategies, achieving complex winning objectives (see Sect.~\ref{subsec:wc} and Sect.~\ref{subsec:bwc}).

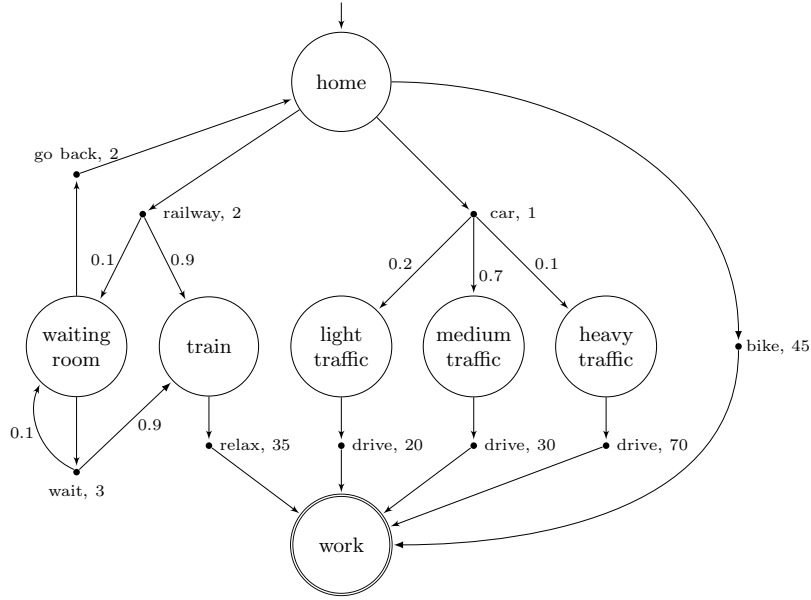
\begin{figure}[t]
  \centering
  \scalebox{0.88}{\begin{tikzpicture}[->,>=latex,shorten >=1pt,auto,node distance=2.5cm,bend angle=45,font=\small, inner sep=2pt]
    \tikzstyle{p1}=[draw,circle,text centered,minimum size=15mm, text width=13mm, inner sep=1pt]
    \tikzstyle{act}=[fill,circle,inner sep=1pt,minimum size=1.5pt, node distance=1cm]
    \tikzstyle{empty}=[text centered, text width=15mm]
    \node[p1] (home) at (0,0) {home};
    \node[p1] (waiting) at (-4,-4) {waiting\\room};
    \node[p1] (train) at (-2,-4) {train};
    \node[p1] (light) at (0,-4) {light\\traffic};
    \node[p1] (medium) at (2,-4) {medium\\traffic};
    \node[p1] (heavy) at (4,-4) {heavy\\traffic};
    \node[p1,double] (work) at (0,-7) {work};
    \node[act] (railway) at (-3,-2) {};
    \node[empty] at ($(railway)+(0.9,0)$) {\scriptsize railway, 2};
    \node[act] (car) at (2,-2) {};
    \node[empty] at ($(car)+(0.6,0)$) {\scriptsize car, 1};
    \node[act] (wait) at (-4,-5.9) {};
    \node[empty] at ($(wait)+(0,-0.3)$) {\scriptsize wait, 3};
    \node[act] (trainAct) at (-2,-5.5) {};
    \node[empty] at ($(trainAct)+(0.7,0)$) {\scriptsize relax, 35};
    \node[act] (back) at (-4,-1.4) {};
    \node[empty] at ($(back)+(0,0.3)$) {\scriptsize go back, 2};
    \node[act] (bike) at (6,-4) {};
    \node[empty] at ($(bike)+(0.6,0)$) {\scriptsize bike, 45};
    \node[act] (lightAct) at (0,-5.5) {};
    \node[empty] at ($(lightAct)+(0.7,0)$) {\scriptsize drive, 20};
    \node[act] (mediumAct) at (2,-5.5) {};
    \node[empty] at ($(mediumAct)+(0.7,0)$) {\scriptsize drive, 30};
    \node[act] (heavyAct) at (4,-5.5) {};
    \node[empty] at ($(heavyAct)+(0.7,0)$) {\scriptsize drive, 70};
    \path[-latex']
    (0,1.2) edge (home)
	(light) edge (lightAct)
	(lightAct) edge (work)
	(mediumAct) edge (work)
	(heavyAct) edge (work)
	(medium) edge (mediumAct)
	(heavy) edge (heavyAct)
    (waiting) edge (back)
    (back) edge (home)
    (train) edge (trainAct)
    (trainAct) edge (work)
    (home) edge (railway)
    (railway) edge  node[left,xshift=0mm]{\scriptsize $0.1$} (waiting)
    (railway) edge  node[right,xshift=0mm]{\scriptsize $0.9$} (train)
    (home) edge (car)
    (car) edge node[left,xshift=-1mm]{\scriptsize $0.2$} (light)
    (car) edge node[right,near end]{\scriptsize $0.7$} (medium)
    (car) edge node[right,xshift=1mm]{\scriptsize $0.1$} (heavy)
    (waiting) edge (wait)
    (wait) edge[bend left] node[left,xshift=0mm]{\scriptsize $0.1$} (waiting)
    ;    
	\draw [->] (home) to[out=0,in=90] (bike);
	\draw [->] (bike) to[out=270,in=0] (work);
	\draw [->] (wait) to node[right,xshift=1mm]{\scriptsize $0.9$}  (train);
  \end{tikzpicture}}
  \caption{An everyday life application of stochastic shortest path problems: choosing a mean of transport to go from home to work. Actions (black dots) are labeled with durations in minutes, and stochastic transitions are labeled with their probability.}
  \label{fig:exampleTS}
\end{figure}

\smallskip\noindent\textbf{Toy example.} A simple example of a \textit{Markov decision process} modeling the rational employee faced to a stochastic environment is depicted in Fig.~\ref{fig:exampleTS}. Vertices model situations in which the employee usually has several choices, represented by outgoing edges. For example, at home, he has the possibility to go to the train station, to take his car or to take his bicycle. His goal is to reach work. Each action has a name, an integer label representing the time taken by the action, and may result in different outcomes according to a discrete probability distribution, modeling the stochastic environment. For example, if the employee decides to take his car ($1$~minute), then the journey depends on traffic conditions and may result\,---\,with respective probability $0.2$, $0.7$ and $0.1$\,---\,in a $20$-minute, $30$-minute or $70$-minute drive. When he decides to bike, he reaches his office in 45 minutes: this action is deterministic, not impacted by the environment. Finally, the employee can also try to catch a train, which takes 35 minutes to reach work. But trains can be delayed (potentially multiple times): in that case, the employee decides if he waits or if he goes back home (and then takes his car or his bike).

We give the formal definition of Markov decision processes and related notions in the following. Informally, Markov decision processes can be seen as two-player games against a stochastic adversary (i.e., the opponent adopts a fixed randomized strategy known to the rational agent). Indeed, they are sometimes called $1\frac{1}{2}$-player games (the half-player being the stochastic one).

\smallskip\noindent\textbf{Markov decision processes.} A \textit{Markov decision process} (MDP) is a tuple $D=(S,s_{\sf init},A,\delta,\weight)$ where
(i) $S$ is a finite set of \emph{states}; (ii) $s_{\sf init} \in S$ is the initial state,; (iii) $A$~is a finite set of \emph{actions}; (iv) $\delta\colon S\times A \rightarrow \calD(S)$
is a partial function called the \emph{probabilistic transition function}, where~$\calD(S)$ denotes the set of rational probability distributions over~$S$; and (v) $w\colon A \rightarrow \mathbb{Z}^d$ is a \emph{$d$-dimensional weight function}.
The set of actions that are available in a state $\state \in \states$ is denoted by $A(s)$. For any dimension~$i \in \{1,\ldots,d\}$, we denote by $\weight_i \colon A \rightarrow \integ$ the projection of~$\weight$ to the~$i$-th dimension, $i$ is omitted when there is only one dimension.

We define a \emph{run} $\rho$ of~$D$ as an infinite sequence $\rho=s_1a_1 \ldots a_{n-1}
s_n\ldots$ of states and actions such that $\delta(s_i,a_i)(s_{i+1})>0$ for all~$i\geq 1$. Essentially, runs in MDPs are analogous to plays in games.

\smallskip\noindent\textbf{Strategies.} As for games, strategies are defined over all valid prefixes of runs (w.r.t.~the underlying graph). Two differences here: first there is only one player in an MDP, the rational agent; second, we allow strategies to use \textit{randomness} in addition to memory. Therefore, a~\emph{strategy}~$\sigma$ is a function $(S A)^{\ast}S \rightarrow \calD(A)$ such that
for all prefix $\eta = s_1a_1 \ldots a_{n-1}
s_n$, we have~$\supp(\sigma(\eta)) \subseteq A(s_n)$, where $\supp$ denotes the support of the probability distribution. As for games, we are interested in the complexity of strategies: whether we need no memory, finite memory or infinite memory; whether we need to use randomness or not.

In order to properly define the different solution concepts proposed in the following, we need to introduce two additional notions: induced Markov chains and the truncated sum payoff function.

\smallskip\noindent\textbf{Markov chains induced by a strategy.} When fixing the strategy $\strat$ of the rational agent in an MDP, the resulting process is fully stochastic and is called a \textit{Markov chain} (MC). We omit technical details here (the interested reader will find exhaustive discussion in~\cite{Puterman-wiley94,filar1997,BK-book08}). Markov chains are essentially MDPs where for all $s \in S$, we have that $\vert A(s)\vert = 1$. Given an MDP $D$ and a strategy $\strat$, let $D^\sigma$ be the induced MC. Assuming~$\strat$ uses only finite memory, then $D^\sigma$ is also finite. 

In this MC $D^\sigma$, an \textit{event} is a measurable set of runs $\event$. Every event has a uniquely defined probability~\cite{Vardi-focs85}: we denote by $\mathbb{P}_{D}^{\strat}(\event)$ the probability that a run of the MDP $D$ belongs to $\event$ when the player follows strategy~$\strat$. Given a measurable function $f$ mapping runs to the numerical domain, we denote by $\expect_{D}^{\strat}(f)$ the \textit{expected value} or \textit{expectation} of $f$ over runs in $D$ induced by strategy $\strat$.

\smallskip\noindent\textbf{Truncated sum payoff.} Let $D=(S,s_{\sf init},A,\delta,\weight)$ be an MDP with a \textit{single-dimensional} weight function $w\colon A \rightarrow \mathbb{N}_{0}$ that assigns to each action $a \in A$ a strictly positive integer. Let $T \subseteq S$ be a set of target states. Given a run $\rho=s_1 s_2 \dots s_i \dots$, with $s_1 = \initState$, in the MDP, we define its \textit{truncated sum} up to $T$ to be $\truncatedSum{T}(\rho) = \sum_{j=1}^{n-1} w(a_j)$ if $s_{n}$ is the first visit of a state in $T \subseteq \states$ within $\rho$; otherwise if $T$ is never reached, then we set $\truncatedSum{T}(\rho) = \infty$. The function $\truncatedSum{T}$ is measurable, and so this function has an expected value in an MC and sets of runs defined from $\truncatedSum{T}$ are measurable.

Intuitively, we consider graphs where all actions induce strictly positive costs and the truncated sum function computes the sum of the weights up to the first visit of a designated target $T$: this is exactly \textit{what we need to model the shortest path problem}. In the following, we survey several solution concepts of interest for this problem.

\subsection{Solution concept $\mathcal{S}_1$: minimizing the expected time-to-work}
\label{subsec:expect}

As mentioned in Sect.~\ref{sec:intro}, the most traditional approach to decision-making in such a stochastic environment is to reason about the \textit{expected payoff}, thus in our case (Fig.~\ref{fig:exampleTS}), to choose the mean of transport that minimizes the expected time-to-work. Such a problem can be formalized as follows.

\begin{definition}[Problem $\mathcal{S}_1$]
Given a single-dimensional weighted MDP $D=(S,s_{\sf init},A,\delta,\weight)$ and a threshold $\ell \in \mathbb{N}$, decide if there exists $\sigma$ such that $\expect^{\sigma}_{D}(\truncatedSum{T}) \leq \ell$.
\end{definition}

This problem is well-studied in the literature, and one can prove that deciding the existence of a strategy achieving an expected time-to-work less than a given threshold can be done in polynomial time, and that such a strategy is simple: it is \textit{pure} (i.e., does not require randomness) and \textit{memoryless} (i.e., it depends only on the current state). In layman's terms, this solution concept is easy to handle and produced strategies are very simple. Formal discussion of this problem can be found in~\cite{bertsekas1991analysis}.

\begin{theorem}[\cite{bertsekas1991analysis}]
\label{thm:SSPE}
Problem $\mathcal{S}_1$ can be decided in polynomial time. Optimal pure memoryless strategies always exist and can be constructed in polynomial time.
\end{theorem}

\renewcommand{\arraystretch}{1.5}
\begin{center}
\begin{tabular}{|>{\arraybackslash}p{5cm}|}
\hline 
$\;$ \underline{\textbf{Solution $\mathcal{S}_1$}}\\
$\qquad$ $\strat$: take the car.\\
$\qquad$ $\expect^{\sigma}_{D}(\truncatedSum{T}) = 33$ minutes.\\
\hline 
\end{tabular} 
\end{center}

In our practical case, it turns out that taking \textit{the car is the strategy that minimizes the expected time-to-work}: this choice gives an expectation equal to 33 minutes. Adopting such a strategy makes sense if one thinks about regular travels, for example when deciding whether to buy a car or a yearly train pass. Indeed, in the long run, the \textit{observed average} time-to-work will be quite close to expectation.

\subsection{Solution concept $\mathcal{S}_2$: traveling without taking too many risks}
\label{subsec:proba}

Observe that taking the car presents some risk: if the traffic is heavy, then work is only reached after $71$ minutes. This can be unacceptable for the employee's boss if it happens too frequently.  So if the employee is {\em risk-averse}, optimizing the expectation may not be the best choice. For example, the employee may want to reach work within $40$ minutes with high probability, say $95\%$. In this case, we need to solve a different problem, called a \textit{percentile problem}. 

\begin{definition}[Problem $\mathcal{S}_2$]
\label{def:sspp}
Given a single-dimensional weighted MDP $D=(S,s_{\sf init},A,\delta,\weight)$, value $\ell \in \mathbb{N}$, and probability threshold $\alpha \in [0,1] \cap \mathbb{Q}$, decide if there exists a strategy $\sigma$ such that $\pr_{D}^\sigma \big[ \{ \rho \mid \truncatedSum{T}(\rho) \leq \ell \} \big] \geq \alpha$.
\end{definition}

This problem is also well-studied in the literature~\cite{Ohtsubo-amc2004,SO-jcta13,DBLP:conf/fossacs/UmmelsB13,HaaseK14,RRS-cav15}. Unfortunately, it is more complex to solve than problem $\mathcal{S}_1$: it can be proved that computing an adequate strategy this objective requires \textit{pseudo-}polynomial time, i.e., exponential time in the length of the binary encoding of the weights. Furthermore, in general the corresponding strategies also require exponential memory.

\begin{theorem}
\label{thm:SSPP}
Problem $\mathcal{S}_2$ can be decided in pseudo-polynomial time, and it is {\sf PSPACE}-hard. Optimal pure strategies with exponential memory always exist and can be constructed in exponential time. 
\end{theorem}

\renewcommand{\arraystretch}{1.5}
\begin{center}
\begin{tabular}{|>{\arraybackslash}p{8cm}|}
\hline 
$\;$ \underline{\textbf{Solution $\mathcal{S}_2$}}\\
$\qquad$ $\strat$: take the train and always wait.\\
$\qquad$ $\pr_{D}^\sigma \big[ \{ \rho \mid \truncatedSum{T}(\rho) \leq \ell \} \big] = 0.99$.\\
\hline 
\end{tabular} 
\end{center}

First, observe that taking the train ensures to reach work within $40$ minutes in $99\%$ of the runs. Indeed, if the train is not delayed, we reach work with $37$ minutes, and this happens with probability $9/10$. Now, if the train is late and the employee decides to wait, the train arrives in the next $3$ minutes with probability $9/10$: in that case, the employee arrives at work within $40$ minutes. So, the strategy consisting in going to the railway station and waiting for the train (as long as needed) gives us a probability $99/100$ to reach work within $40$ minutes, fulfilling our objective.

Second, it is easy to see that both bicycle and car are excluded in order to satisfy the problem $\mathcal{S}_2$. With the bicycle we reach work in $45$ minutes with probability one, and with the car we reach work in $71$ minutes with probability $1/10$, hence we miss the constraint of $40$ minutes too often.   

\subsection{Solution concept $\mathcal{S}_3$: strict worst-case guarantees}
\label{subsec:wc}

Assume now that the employee wants a strategy to go from home to work such that work is \textit{guaranteed} to be reached within 60 minutes (e.g., to avoid missing an important meeting with his boss). It is clear that both previously defined optimal (w.r.t.~problems $\mathcal{S}_1$ and $\mathcal{S}_2$ respectively) strategies are excluded: there is the possibility of heavy traffic with the car (and a journey of $71$ minutes), and trains can be delayed indefinitely in the \textit{worst case}. While the probability of such an event is actually zero, the mere existence of the corresponding run is still not acceptable when reasoning about worst-case behaviors. Moreover, whatever the threshold fixed on the acceptable time-to-work, the train option will exceed it with positive\,---\,yet very small\,---\,probability.

To ensure a strict upper bound on the reaching-time, an adequate model is to bring back \textit{rationality} to the environment and to consider it, not anymore as a stochastic agent, but as an \textit{antagonistic opponent}. The result is a two-player zero-sum game as in Sect.~\ref{sec:synthesis}, where the uncertainty becomes adversarial: when there is some uncertainty about the outcome of an action, we do not consider a probabilistic model but we let an {\em adversary} decide the outcome of the action. The problem is defined as follows.

\begin{definition}[Problem $\mathcal{S}_3$]
Given single-dimensional weighted MDP $D = (S,s_{\sf init},A,\delta,\weight)$, set of target states $T \subseteq S$, and value $\ell \in \mathbb{N}$, decide if there exists a strategy $\sigma$ such that for all run $\rho$, we have that $\truncatedSum{T}(\rho) \leq \ell$.
\end{definition}

In this case, we again find an easy setting: polynomial time suffices to decide if an adequate strategy exists, and such strategies require neither memory nor randomness. More details can be found in~\cite{BrihayeGHM14,FiliotGR12,Chatterjee201525}.

\begin{theorem}[\cite{DBLP:dblp_journals/mst/KhachiyanBBEGRZ08}]
Problem $\mathcal{S}_3$ can be decided in polynomial time. Optimal pure memoryless strategies always exist and can be constructed in polynomial time. 
\end{theorem}

\renewcommand{\arraystretch}{1.5}
\begin{center}
\begin{tabular}{|>{\arraybackslash}p{8cm}|}
\hline 
$\;$ \underline{\textbf{Solution $\mathcal{S}_3$}}\\
$\qquad$ $\strat$: take the bicycle.\\
$\qquad$ Worst-case reaching-time: 45 minutes.\\
\hline 
\end{tabular} 
\end{center}

If we apply this technique to our running example, we obtain that taking the bicycle is a safe option to ensure the strict $60$ minutes upper bound. Indeed, its worst-case reaching-time is 45 minutes. However, the expected time-to-work when following this strategy is also $45$ minutes, which is far from the optimum of $33$ minutes that can be obtained when we neglect the worst-case constraint (solution $\mathcal{S}_1$). The goal of the next solution concept is exactly to find a strategy allying the best aspects of solutions $\mathcal{S}_1$ and $\mathcal{S}_3$.

\subsection{Solution concept $\mathcal{S}_4$: minimizing the expected time under strict worst-case guarantees}
\label{subsec:bwc}

In answer to the previous situation, the employee may be interested in synthesizing a strategy that \textit{minimizes the expected time-to-work under the constraint that work is reached within 60 minutes in the worst case}. This is a complex objective which cannot be expressed in the traditional frameworks. Intuitively, it combines two different views of the environment: for the expected value, the environment is seen as a \textit{stochastic agent}, while for the worst case, it is seen as a \textit{rational antagonistic agent}. In that sense, it constitutes an example of what we call \textit{rich behavioral models}.

In recent joint work~\cite{DBLP:journals/corr/BruyereFRR14,bruyere_STACS2014}, we developed sound formal grounds for the analysis of such complex objectives. We called the corresponding problem the \textit{beyond worst-case synthesis problem}. Its goal is to look for strategies that ensure, \textit{simultaneously}, a worst-case threshold (when probabilities are replaced by adversarial choices), and a good expectation (when probabilities are taken into account). It is formally defined as follows.

\begin{definition}[Problem $\mathcal{S}_4$]
Given a single-dimensional weighted MDP $D=(S,s_{\sf init},A,\delta,\weight)$, a set of target states $T \subseteq S$, and two values $\ell_1,\ell_2 \in \mathbb{N}$, decide if there exists a strategy $\sigma$ such that:
  \begin{enumerate}
  	\item For all induced run $\rho$, $\truncatedSum{T}(\rho) \leq \ell_1$,
  	\vspace{1mm}
	\item $\expect^{\sigma}_{D}(\truncatedSum{T}) \leq \ell_2$.
  \end{enumerate}
\end{definition}

This problem can be seen as a generalization subsuming both problems $\mathcal{S}_1$ and $\mathcal{S}_3$. While those problems are both solvable in polynomial time and pure memoryless strategies suffice in both cases, problem $\mathcal{S}_4$ proves to be inherently harder. We showed that, similarly to problem $\mathcal{S}_2$, pseudo-polynomial time is necessary to construct winning strategies and those strategies require memory (but no randomness). Extensive discussion of this novel framework and ad-hoc techniques can be found in~\cite{DBLP:journals/corr/BruyereFRR14,bruyere_STACS2014}.

\begin{theorem}[\cite{bruyere_STACS2014}]
Problem $\mathcal{S}_4$ can be decided in pseudo-poly\-no\-mial time and is {\sf NP}-hard. Pseudo-poly\-nomial memory is always sufficient and in general necessary, and satisfying strategies can be constructed in pseudo-polynomial time. 
\end{theorem}

\renewcommand{\arraystretch}{1.5}
\begin{center}
\begin{tabular}{|>{\arraybackslash}p{10.5cm}|}
\hline 
$\;$ \underline{\textbf{Solution $\mathcal{S}_4$}}\\
$\qquad$ $\strat$: try to take the train, if the train is delayed three times\newline \hspace*{12.5mm}consecutively, then go back home and take the bicycle.\\
$\qquad$ $\expect^{\sigma}_{D}(\truncatedSum{T}) \approx 37.34$ minutes.\\
$\qquad$ Worst-case reaching-time: 58 minutes.\\
\hline 
\end{tabular} 
\end{center}

For our employee, the optimal strategy w.r.t.~problem $\mathcal{S}_4$ is the following: try to take the train, if the train is delayed three times consecutively, then go back home and take the bicycle. This strategy is safe as it always reaches work within 58 minutes and its expectation is $\approx 37.34$ minutes (much better than taking directly the bicycle). Hence, this a perfectly suitable choice for the employee. However, it comes at a cost in terms of strategy complexity: observe that this strategy requires finite memory (because the employee must be able to recall how many times the train has been delayed already), in contrast to solutions $\mathcal{S}_1$ and $\mathcal{S}_3$.

\subsection{Solution concept $\mathcal{S}_5$: trade-offs between multiple objectives}
\label{subsec:percentile}

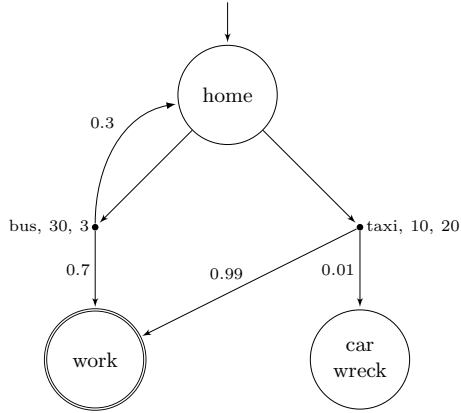
\begin{figure}[htb]
  \centering
  \scalebox{0.88}{\begin{tikzpicture}[->,>=latex,shorten >=1pt,auto,node distance=2.5cm,bend angle=45,font=\small, inner sep=2pt]
    \tikzstyle{p1}=[draw,circle,text centered,minimum size=15mm, text width=13mm, inner sep=1pt]
    \tikzstyle{act}=[fill,circle,inner sep=1pt,minimum size=1.5pt, node distance=1cm]
    \tikzstyle{empty}=[text centered, text width=15mm]
    \node[p1] (home) at (0,0) {home};
    \node[p1,double] (work) at (-2,-4) {work};
    \node[p1] (wreck) at (2,-4) {car\\wreck};
    \node[act] (bus) at (-2,-2) {};
    \node[empty] at ($(bus)+(-0.7,0)$) {\scriptsize bus, 30, 3};
    \node[act] (taxi) at (2,-2) {};
    \node[empty] at ($(taxi)+(0.8,0)$) {\scriptsize taxi, 10, 20};
    \path[-latex']
    (0,1.4) edge (home)
    (home) edge (bus)
    (bus) edge node[left,xshift=0mm,yshift=0mm]{\scriptsize $0.7$} (work)
    (home) edge (taxi)
    (taxi) edge node[left,xshift=0mm,yshift=1.4mm]{\scriptsize $0.99$} (work)
    (taxi) edge node[left,xshift=0mm,yshift=0mm]{\scriptsize $0.01$} (wreck)
    ;    
	\draw [->] (bus) to[out=90,in=190] node[above,xshift=-2mm,yshift=2mm]{\scriptsize $0.3$} (home);
  \end{tikzpicture}}
  \caption{Multi-constraint percentile queries can help when actions both impact the duration of the journey (first dimension) and its cost (second dimension): trade-offs have to be considered.}
  \label{fig:percentile}
\end{figure}

The last solution concept we describe will be illustrated on a different example. The MDP on Fig.~\ref{fig:percentile} represents a simplified choice model for commuting, but introduces two-dimensional weights: each action is labeled with a duration, in minutes, and a cost, in dollars. Multi-dimensional MDPs are useful to analyze systems with {\em multiple objectives} that are potentially conflicting and make necessary the analysis of trade-offs. For instance, we may want a choice of transportation that gives us high probability to reach work in due time but also limits the risk of an expensive journey. Since faster options are often more expensive, \textit{trade-offs} have to be considered.

Recall problem $\mathcal{S}_2$: it asks to decide the existence of strategies satisfying a \textit{single percentile constraint}. As all models discussed up to now, this problem can only be applied to single-dimensional MDPs. For example, one may look for a strategy that ensures that $80\%$ of the runs take at most $40$ minutes (constraint C1), \textit{or} that $50\%$ of them cost at most $10$ dollars (C2). A good strategy for C1 would be to take the taxi, which guarantees that work is reached within $10$ minutes with probability $0.99 > 0.8$. For C2, taking the bus is a good option, because already $70\%$ of the runs will reach work for only $3$ dollars. Note that taking the taxi does not satisfy C2, nor does taking the bus satisfy C1.

In practice, a desirable strategy should be able to satisfy \textit{both} C1 and C2. This is the goal of our model of \textit{multi-constraint percentile queries}, introduced in~\cite{RRS-cav15}. It can be formalized as follows.

\begin{definition}[Problem $\mathcal{S}_5$]
Given a $d$-dimensional weighted MDP $D=(S,s_{\sf init},A,\delta,\weight)$, and $q \in \nat$ percentile constraints described by sets of target states $T_{i} \subseteq S$, dimensions $k_{i} \in \{1, \ldots{}, d\}$, value thresholds $\ell_{i} \in \nat$ and probability thresholds $\alpha_{i} \in [0,1] \cap \rat$, where $i \in \{1, \ldots{}, q\}$, decide if there exists a strategy $\sigma$ such that
\begin{equation*}
\forall\, i\in \{1, \ldots{}, q\},\; \pr_{D}^\sigma \big[ \{ \rho \mid \truncatedSum{\truncatedTarget_{i}}_{k_{i}}(\rho) \leq \ell_i \} \big] \geq \alpha_i
\end{equation*}
where $\truncatedSum{\truncatedTarget_{i}}_{k_{i}}$ denotes the truncated sum on dimension $k_{i}$ and w.r.t.~target set $\truncatedTarget_{i}$.
\end{definition}

Our technique is able to solve the problem for queries with multiple constraints, potentially related to different dimensions of the weight function and to different target sets: this offers great flexibility which is useful in modeling practical applications. In terms of complexity, two things should be noted. First, while significantly more expressive than problem $\mathcal{S}_2$, this problem does not induce a blow-up in complexity: the algorithmic complexity and the memory needs for strategies are similar. Second, among the solution concepts presented in this paper, problem $\mathcal{S}_5$ is the only one that actually requires strategies to use randomness in full generality.

\begin{theorem}[\cite{RRS-cav15}]
Problem $\mathcal{S}_5$ can be decided in exponential time in general, and pseudo-polynomial time for single-dimension single-target multi-constraint queries. The problem is {\sf PSPACE}-hard even for single-constraint que\-ries. Randomized exponential-memory strategies are always sufficient and in general necessary, and satisfying strategies can be constructed in exponential time.
\end{theorem}

\renewcommand{\arraystretch}{1.5}
\begin{center}
\begin{tabular}{|>{\arraybackslash}p{11.5cm}|}
\hline 
$\;$ \underline{\textbf{Solution $\mathcal{S}_5$}}\\
$\qquad$ $\strat$: try the bus once, then take the taxi if the bus does not depart.\\
$\qquad$ $\pr_{D}^\sigma \big[ \{ \rho \mid \truncatedSum{T}_{1}(\rho) \leq 40 \} \big] > 0.99$.\\
$\qquad$ $\pr_{D}^\sigma \big[ \{ \rho \mid \truncatedSum{T}_{2}(\rho) \leq 10 \} \big] = 0.7$.\\
\hline 
\end{tabular} 
\end{center}

Let us see what can be an appropriate strategy for the conjunction (C1 $\wedge$ C2) in our example. One is to try the bus once, and then take the taxi if the bus does not depart. Indeed, this strategy ensures that work is reached within $40$ minutes with probability larger than $0.99$ thanks to runs home$\cdot$bus$\cdot$work (probability $0.7$ and duration $30$) and home$\cdot$bus$\cdot$home$\cdot$taxi$\cdot$work (probability $0.297$ and duration $40$). Furthermore, it also ensures that more than half the time, the total cost to target is at most $10$ dollars, thanks to run home$\cdot$bus$\cdot$work which has probability $0.7$ and cost $3$. Observe that this strategy requires \textit{memory}.

\renewcommand{\arraystretch}{1.5}
\begin{center}
\begin{tabular}{|>{\arraybackslash}p{11.5cm}|}
\hline 
$\;$ \underline{\textbf{Solution $\mathcal{S}_5$}}\\
$\qquad$ $\strat'$: take the bus (resp.~the taxi) with probability $3/5$ (resp.~$2/5$).\\
$\qquad$ $\pr_{D}^\sigma \big[ \{ \rho \mid \truncatedSum{T}_{1}(\rho) \leq 40 \} \big] > 0.81$.\\
$\qquad$ $\pr_{D}^\sigma \big[ \{ \rho \mid \truncatedSum{T}_{2}(\rho) \leq 10 \} \big] > 0.5$.\\
\hline 
\end{tabular} 
\end{center}

In this particular example, it is possible to build another acceptable strategy which is memoryless but requires \textit{randomness}. Consider the strategy that flips an unfair coin in home to decide if we take the bus or the taxi, with probabilities $3/5$ and $2/5$ respectively. Constraint C1 is ensured thanks to runs home$\cdot$bus$\cdot$work (probability $0.42$) and home$\cdot$taxi$\cdot$work (probability $0.396$). Constraint C2 is ensured thanks to runs (home$\cdot$bus)$^{n}\cdot$work with $n = 1, 2, 3$: they have probabilities $0.42$, $\geq 0.07$ and $\geq 0.01$ respectively, totaling to $\geq 0.5$, while they all have cost at most $3\cdot 3 = 9 < 10$.

\section{Conclusion}
\label{sec:conclusion}

Through this contribution, we discussed two traditional paradigms used for describing the behavior of agents in complex systems: \textit{rationality} and \textit{stochasticity}. Both have proved to be powerful abstraction mechanisms in many contexts. The main point of this paper is that it is sometimes necessary to bring together these two paradigms in \textit{rich behavioral models}.

Our core research field is \textit{controller synthesis} for reactive systems, hence the models we develop often arise from problems linked to computer systems. In Sect.~\ref{sec:synthesis}, we gave an overview of the application of game theory for controller synthesis, highlighting the crux of the approach and the main challenges.

In Sect.~\ref{sec:rich}, we surveyed five solution concepts, illustrated on an everyday life example: how to plan a journey in an uncertain environment. We presented three classical approaches, based on the traditional dichotomy between rational and stochastic environments, and then discussed recent advances breaching this dichotomy. The simple examples presented in the paper are already sufficient to witness that such rich behavioral models do have great practical interest.

Our presentation was at a very high level, aimed at a broad audience, but the reader interested in actual mathematical models, algorithms and tools will hopefully find more information in the corresponding cited papers. It is also worth noting that all concepts presented here for the shortest path are indeed applicable in various, more general frameworks (e.g., many other payoff functions have been studied).

While our work is motivated by applications to controller synthesis, we are led to believe that such models can also prove interesting in other areas. For example, related models have been used in economics~\cite{DBLP:conf/fsttcs/BergerKSV08} to model investor profiles as strategies in games with stochastic aspects. With that in mind, beyond worst-case strategies (solution concept $\mathcal{S}_4$) that ensure both sufficient risk-avoidance and profitable expected return can be of interest, as well as related models of value-at-risk~\cite{RRS-cav15,RRS15b}.

\bibliographystyle{plain}
\bibliography{biblio}

\begin{thebibliography}{10}

\bibitem{BK-book08}
C.~Baier and J.-P. Katoen.
\newblock {\em Principles of model checking}.
\newblock MIT Press, 2008.

\bibitem{DBLP:conf/fsttcs/BergerKSV08}
N.~Berger, N.~Kapur, L.J. Schulman, and V.V. Vazirani.
\newblock Solvency games.
\newblock In {\em Proc. of FSTTCS}, LIPIcs 2, pages 61--72. Schloss Dagstuhl -
  LZI, 2008.

\bibitem{bertsekas1991analysis}
D.P. Bertsekas and J.N. Tsitsiklis.
\newblock An analysis of stochastic shortest path problems.
\newblock {\em Mathematics of Operations Research}, 16(3):580--595, 1991.

\bibitem{DBLP:conf/cav/BloemCHJ09}
R.~Bloem, K.~Chatterjee, T.A. Henzinger, and B.~Jobstmann.
\newblock Better quality in synthesis through quantitative objectives.
\newblock In {\em Proc. of CAV}, LNCS 5643, pages 140--156. Springer, 2009.

\bibitem{DBLP:conf/tacas/BohyBFR13}
A.~Bohy, V.~Bruy{\`e}re, E.~Filiot, and J.-F. Raskin.
\newblock Synthesis from {LTL} specifications with mean-payoff objectives.
\newblock In {\em Proc. of TACAS}, LNCS 7795, pages 169--184. Springer, 2013.

\bibitem{DBLP:journals/corr/BouyerMRLL15}
P.~Bouyer, N.~Markey, M.~Randour, K.G. Larsen, and S.~Laursen.
\newblock Average-energy games.
\newblock In {\em Proc. of GandALF}, {EPTCS} 193, pages 1--15, 2015.

\bibitem{LATA}
R.~Brenguier, L.~Clemente, P.~Hunter, {G.A.} P\'erez, M.~Randour, J.-F. Raskin,
  O.~Sankur, and M.~Sassolas.
\newblock Non-zero sum games for reactive synthesis.
\newblock In {\em Proc. of LATA}, LNCS 9618, pages 3--23. Springer, 2016.

\bibitem{BrihayeGHM14}
T.~Brihaye, G.~Geeraerts, A.~Haddad, and B.~Monmege.
\newblock To reach or not to reach? {E}fficient algorithms for total-payoff
  games.
\newblock In {\em Proc. of CONCUR}, LIPIcs 42, pages 297--310. Schloss Dagstuhl
  - LZI, 2015.

\bibitem{DBLP:journals/corr/BruyereFRR14}
V.~Bruy{\`{e}}re, E.~Filiot, M.~Randour, and J.-F. Raskin.
\newblock Expectations or guarantees? {I} want it all! {A} crossroad between
  games and {MDP}s.
\newblock In {\em Proc. of SR}, {EPTCS} 146, pages 1--8, 2014.

\bibitem{bruyere_STACS2014}
V.~Bruy{\`{e}}re, E.~Filiot, M.~Randour, and J.-F. Raskin.
\newblock Meet your expectations with guarantees: Beyond worst-case synthesis
  in quantitative games.
\newblock In {\em Proc. of STACS}, LIPIcs 25, pages 199--213. Schloss Dagstuhl
  - LZI, 2014.

\bibitem{DBLP:conf/emsoft/ChakrabartiAHS03}
A.~Chakrabarti, L.~de~Alfaro, T.A. Henzinger, and M.~Stoelinga.
\newblock Resource interfaces.
\newblock In {\em Proc. of EMSOFT}, LNCS 2855, pages 117--133. Springer, 2003.

\bibitem{Chatterjee201525}
K.~Chatterjee, L.~Doyen, M.~Randour, and J.-F. Raskin.
\newblock Looking at mean-payoff and total-payoff through windows.
\newblock {\em Information and Computation}, 242:25--52, 2015.

\bibitem{DBLP:journals/acta/ChatterjeeRR14}
K.~Chatterjee, M.~Randour, and J.-F. Raskin.
\newblock Strategy synthesis for multi-dimensional quantitative objectives.
\newblock {\em Acta Informatica}, 51(3-4):129--163, 2014.

\bibitem{cherkassky1996shortest}
B.V. Cherkassky, A.V. Goldberg, and T.~Radzik.
\newblock Shortest paths algorithms: Theory and experimental evaluation.
\newblock {\em Math. programming}, 73(2):129--174, 1996.

\bibitem{church1957applications}
A.~Church.
\newblock Applications of recursive arithmetic to the problem of circuit
  synthesis.
\newblock {\em Summaries of the Summer Institute of Symbolic Logic}, 1:3--50,
  1957.

\bibitem{DBLP:conf/lop/ClarkeE81}
E.M. Clarke and E.A. Emerson.
\newblock Design and synthesis of synchronization skeletons using
  branching-time temporal logic.
\newblock In {\em Logic of Programs}, LNCS 131, pages 52--71. Springer, 1981.

\bibitem{em79}
A.~Ehrenfeucht and J.~Mycielski.
\newblock Positional strategies for mean payoff games.
\newblock {\em International Journal of Game Theory}, 8:109--113, 1979.

\bibitem{filar1997}
J.~Filar and K.~Vrieze.
\newblock {\em Competitive {M}arkov decision processes}.
\newblock Springer, 1997.

\bibitem{FiliotGR12}
E.~Filiot, R.~Gentilini, and J.-F. Raskin.
\newblock Quantitative languages defined by functional automata.
\newblock In {\em Proc. of CONCUR}, LNCS 7454, pages 132--146. Springer, 2012.

\bibitem{DBLP:conf/dagstuhl/2001automata}
E.~Gr{\"a}del, W.~Thomas, and T.~Wilke, editors.
\newblock {\em Automata, Logics, and Infinite Games: A Guide to Current
  Research}, LNCS 2500. Springer, 2002.

\bibitem{HaaseK14}
C.~Haase and S.~Kiefer.
\newblock The odds of staying on budget.
\newblock In {\em Proc. of ICALP}, LNCS 9135, pages 234--246. Springer, 2015.

\bibitem{DBLP:dblp_journals/mst/KhachiyanBBEGRZ08}
L.~Khachiyan, E.~Boros, K.~Borys, K.M. Elbassioni, V.~Gurvich, G.~Rudolf, and
  J.~Zhao.
\newblock On short paths interdiction problems: Total and node-wise limited
  interdiction.
\newblock pages 204--233, 2008.

\bibitem{Ohtsubo-amc2004}
Y.~Ohtsubo.
\newblock Optimal threshold probability in undiscounted markov decision
  processes with a target set.
\newblock {\em Applied Math. and Computation}, 149(2):519 -- 532, 2004.

\bibitem{osborne1994course}
M.J. Osborne and A.~Rubinstein.
\newblock {\em A Course in Game Theory}.
\newblock MIT Press, 1994.

\bibitem{DBLP:conf/popl/PnueliR89}
A.~Pnueli and R.~Rosner.
\newblock On the synthesis of a reactive module.
\newblock In {\em Proc. of POPL}, pages 179--190. ACM Press, 1989.

\bibitem{Puterman-wiley94}
M.L. Puterman.
\newblock {\em Markov Decision Processes: Discrete Stochastic Dynamic
  Programming}.
\newblock John Wiley \& Sons, Inc., New York, NY, USA, 1st edition, 1994.

\bibitem{ramadge1987supervisory}
P.J. Ramadge and W.M. Wonham.
\newblock Supervisory control of a class of discrete event processes.
\newblock {\em SIAM journal on control and optimization}, 25(1):206--230, 1987.

\bibitem{Ran13}
M.~Randour.
\newblock Automated synthesis of reliable and efficient systems through game
  theory: A case study.
\newblock In {\em Proceedings of the European Conference on Complex Systems
  2012}, Springer Proceedings in Complexity XVII, pages 731--738. Springer,
  2013.

\bibitem{randour2014Thesis}
M.~Randour.
\newblock {\em Synthesis in Multi-Criteria Quantitative Games}.
\newblock PhD thesis, UMONS, Universit\'e de Mons, Belgium, 2014.

\bibitem{RRS-cav15}
M.~Randour, J.-F. Raskin, and O.~Sankur.
\newblock Percentile queries in multi-dimensional {M}arkov decision processes.
\newblock In {\em Proc. of CAV}, LNCS 9206, pages 123--139. Springer, 2015.

\bibitem{RRS15b}
M.~Randour, J.-F. Raskin, and O.~Sankur.
\newblock Variations on the stochastic shortest path problem.
\newblock In {\em Proc. of VMCAI}, LNCS 8931, pages 1--18. Springer, 2015.

\bibitem{SO-jcta13}
M.~Sakaguchi and Y.~Ohtsubo.
\newblock Markov decision processes associated with two threshold probability
  criteria.
\newblock {\em Journal of Control Theory and Applications}, 11(4):548--557,
  2013.

\bibitem{DBLP:conf/fossacs/UmmelsB13}
M.~Ummels and C.~Baier.
\newblock Computing quantiles in {M}arkov reward models.
\newblock In {\em Proc. of FOSSACS}, LNCS 7794, pages 353--368. Springer, 2013.

\bibitem{Vardi-focs85}
M.Y. Vardi.
\newblock Automatic verification of probabilistic concurrent finite state
  programs.
\newblock In {\em Proc. of FOCS}, pages 327--338. IEEE Computer Society, 1985.

\bibitem{DBLP:conf/lics/VardiW86}
M.Y. Vardi and P.~Wolper.
\newblock An automata-theoretic approach to automatic program verification.
\newblock In {\em Proc. of LICS}, pages 332--344. IEEE Computer Society, 1986.

\bibitem{DBLP:journals/iandc/VelnerC0HRR15}
Y.~Velner, K.~Chatterjee, L.~Doyen, T.A. Henzinger, A.M. Rabinovich, and J.-F.
  Raskin.
\newblock The complexity of multi-mean-payoff and multi-energy games.
\newblock {\em Information and Computation}, 241:177--196, 2015.

\bibitem{von1944theory}
J.~von Neumann and O.~Morgenstern.
\newblock {\em Theory of Games and Economic Behavior}.
\newblock Princeton University Press, 1944.

\bibitem{ZP96}
U.~Zwick and M.~Paterson.
\newblock The complexity of mean payoff games on graphs.
\newblock {\em Theoretical Computer Science}, 158:343--359, 1996.

\end{thebibliography}
\end{document}